\documentclass[amssymb, preprintnumbers, showpacs, showkeys, aps, prl, superscriptaddress, twocolumn, longbibliography]{revtex4-2}

\usepackage[hidelinks, hypertexnames=false]{hyperref}

\usepackage[all]{hypcap}
\usepackage{xcolor}
\usepackage{slashed}
\usepackage{graphicx}
\usepackage{amsmath}
\usepackage{latexsym}
\usepackage{epstopdf}
\usepackage{amsmath}
\usepackage{enumitem}
\usepackage{amssymb,amsmath}
\usepackage{multirow}
\usepackage{mathtools}
\usepackage{bbm}
\usepackage{bm}
\usepackage{amsfonts}
\usepackage{amssymb}
\usepackage{calligra}
\usepackage{calrsfs}
\usepackage{makecell}
\usepackage[normalem]{ulem}
\usepackage[USenglish,american]{babel}
\usepackage{physics}
\usepackage{braket}

\newcommand{\be}{\begin{equation}}
\newcommand{\ee}{\end{equation}}
\newcommand{\ba}{\begin{eqnarray}}
\newcommand{\ea}{\end{eqnarray}}
\newcommand{\bs}{\begin{split}}
\newcommand{\es}{\end{split}}

\begin{document}

\title{Supersolid phases of bosonic particles in a bubble trap}

\author{Matteo Ciardi}
\email{matteo.ciardi@tuwien.ac.at}
\affiliation{Dipartimento di Fisica e Astronomia, Universit\`a di Firenze, I-50019, Sesto Fiorentino (FI), Italy}
\affiliation{INFN, Sezione di Firenze, I-50019, Sesto Fiorentino (FI), Italy}
\affiliation{Institute for Theoretical Physics, TU Wien, Wiedner Hauptstraße 8-10/136, 1040 Vienna, Austria}

\author{Fabio Cinti}
\email{fabio.cinti@unifi.it}
\affiliation{Dipartimento di Fisica e Astronomia, Universit\`a di Firenze, I-50019, Sesto Fiorentino (FI), Italy}
\affiliation{INFN, Sezione di Firenze, I-50019, Sesto Fiorentino (FI), Italy}
\affiliation{Department of Physics, University of Johannesburg, P.O. Box 524, Auckland Park 2006, South Africa}

\author{Giuseppe Pellicane}
\email{giuseppe.pellicane@unime.it}
\affiliation{Dipartimento di Scienze Biomediche, Odontoiatriche e delle Immagini Morfologiche e Funzionali, Universit\`a degli Studi di Messina, I-98125, Messina, Italy}
\affiliation{CNR-IPCF, Viale F. Stagno d'Alcontres, 37-98158, Messina, Italy}
\affiliation{School of Chemistry and Physics, University of Kwazulu-Natal, 3209 Pietermaritzburg, South Africa}
\affiliation{National Institute of Theoretical and Computational Sciences (NIThECS), 3209 Pietermaritzburg, South Africa}

\author{Santi Prestipino}
\email{sprestipino@unime.it}
\affiliation{Dipartimento di Scienze Matematiche e Informatiche, Scienze Fisiche e Scienze della Terra, Universit\`a degli Studi di Messina, viale F. Stagno d'Alcontres 31, I-98166, Messina, Italy}

\begin{abstract}
Confinement can have a considerable effect on the behavior of particle systems, and is therefore an effective way to discover new phenomena. A notable example is a system of identical bosons at low temperature under an external field mimicking an isotropic bubble trap, which constrains the particles to a portion of space close to a spherical surface. Using Path Integral Monte Carlo simulations, we examine the spatial structure and superfluid fraction in two emblematic cases. First, we look at soft-core bosons, finding the existence of supersolid cluster arrangements with polyhedral symmetry; we show how different numbers of clusters are stabilized depending on the trap radius and the particle mass, and we characterize the temperature behavior of the cluster phases. A detailed comparison with the behavior of classical soft-core particles is provided too. Then, we examine the case, of more immediate experimental interest, of a dipolar condensate on the sphere, demonstrating how a quasi-one-dimensional supersolid of clusters is formed on a great circle for realistic values of density and interaction parameters. Crucially, this supersolid phase is only slightly disturbed by gravity. We argue that the predicted phases can be revealed in magnetic traps with spherical-shell geometry, possibly even in a lab on Earth. Our results pave the way for future simulation studies of correlated quantum systems in curved geometries.
\end{abstract}

\maketitle

Computing the equilibrium properties of quantum many-body systems remains a main objective of contemporary physics. In this respect, ultracold atoms provide a framework where geometry and interactions can be tuned almost at will, allowing to test fundamental many-body theories~\cite{RevModPhys.80.885,Bloch:2012fk}. A system of atoms (loosely) confined to an ellipsoidal surface~\cite{Garraway2016,Perrin2017} represents a typology that only recently has started to be explored. To achieve this goal, a quantum gas is loaded into a shell trap, where atoms are subject to a quadrupolar field ``dressed'' by a radiofrequency (rf) field~\cite{PhysRevLett.86.1195,PhysRevA.69.023605, Colombe2004, Morizot2007, Harte2018}. In the limit of slow atomic motion in a strong magnetic field, the effective potentials of the internal states are the position-dependent eigenvalues of the Hamiltonian consisting of the bare potentials and the coupling term~\cite{Messiah_2014a}. For atoms in the upper dressed state, resonance is reached at the surface of an ellipsoid. However, to let atoms explore the full surface, experiments must be performed in outer space~\cite{Lundblad2019,Aveline2020,Carollo2022} or adopt some gravity compensation mechanism~\cite{Guo2022,PhysRevLett.129.243402}. Coherent and isotropic shells of atoms slowly expanding in microgravity can be generated too~\cite{Meister2019}.

\begin{figure}[t!]
  \begin{center}
    \includegraphics[width=\linewidth]{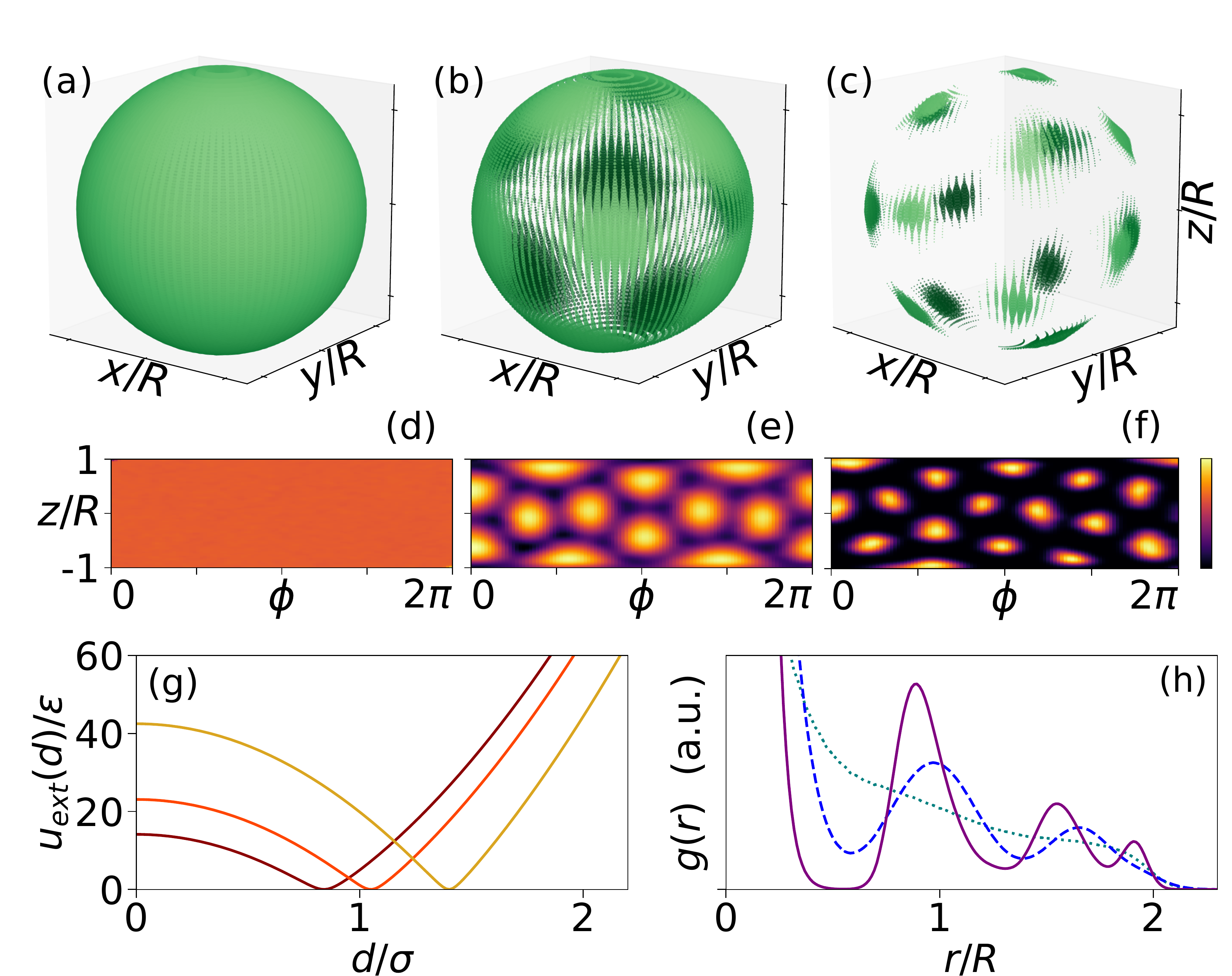}
    \caption{Soft-core bosons in an isotropic bubble trap. (a--c): Particle density for $N=120,T=0.125,R=1.4,u_0=2,\Omega=0.0441$, and $\lambda=0.5$ (a), 0.16 (b), 0.01 (c). The size of points is proportional to the density averaged along the radial direction and the shading is a guide for the eye: brighter points are closer to the observer. (d--f): Area-preserving projections of the densities in (a--c). Brighter colors indicate a larger density (log scale). (g): External potential \eqref{eq2} for $u_0 = 2$, $\Omega=0.0441$, and $R=0.84$ (brown), 1.05 (red), and 1.4 (gold). (h): Pair distribution function for the same parameters in (a--c): $\lambda=0.5$ (blue dotted line), 0.16 (blue dashed line), 0.01 (purple solid line).}
    \label{fig1}
  \end{center}
\end{figure} 

The realization of shell-shaped condensates has fueled a renewed interest in the problem of quantum particles in curved geometries~\cite{Moller2020}, and particularly in the quantum phases that free and interacting bosons can exhibit.
Several recent works have investigated Bose-Einstein condensation as well as the superfluid-to-normal fluid transition on a sphere~\cite{PhysRevLett.123.160403,Bereta2019,PhysRevLett.125.010402,Tononi2022,PhysRevA.75.013611,PhysRevLett.125.010402}, while others have studied the dynamics and thermodynamics of the condensate itself~\cite{Padavic2017,PhysRevA.98.013609,PhysRevA.104.063310,Diniz2020}. 
Most of these works have considered weakly-interacting particles, although some have looked into the condensed phases arising from dipolar interactions~\cite{adhikari2012,Diniz2020,PhysRevA.96.013627,arazo21}.

While experiments have so far been performed in the same dilute limit, bubble traps open the exciting prospect of investigating the physics of strongly correlated quantum particles in curved spaces, with all the advantages brought by ultracold-atom setups. For example, ultracold atoms may serve as quantum simulators to test fundamental physics; therefore, studying the effects of curvature in a controlled environment can be of interest to other fields, ranging from cosmology to biology \cite{PhysRevX.8.021021,RevModPhys.93.025008}. Similarly, we can envisage the possibility to control the self-assembly of a many-body system through a convenient choice of the confining surface, i.e., of the shape of the ensuing geometric potential~\cite{Moller2020}.
Equally important is to figure out experiments on curved many-body systems that could be accomplished on Earth, i.e., without the necessity of compensating gravity~\cite{Lundblad_2023}.
In this respect, the most awaited developments concern dipolar atoms~\cite{Lahaye2009}, which have already been examined in flat space and harmonic traps~\cite{PhysRevX.12.021019, Chomaz2022}, and are known to give rise to supersolid phases~\cite{Norcia2021,Tanzi2021,Chomaz2019,Bottcher2019}. At the same time, it will be interesting to see whether the confined geometry of bubble traps can stabilize the supersolid phase in small systems of Rydberg-dressed atoms~\cite{Browaeys2020,henkel10,PhysRevLett.104.223002,Cinti2010b,Balewski2014}.

In this letter, we use the Path integral Monte Carlo (PIMC) simulation method to give a first glimpse of the equilibrium phases that can arise in a shell-trapped system of identical bosonic particles. In particular, we will provide the first compelling evidence of supersolid order for two distinct instances of the interaction potential. First, we look at the penetrable-sphere model as an example of soft-core potential and a paradigmatic interaction in condensed matter physics that gives rise to both superfluidity and clusterization~\cite{PhysRevA.88.033618,Cinti:2014aa,PhysRevB.86.060510,prestipino18}; for this interaction, we compare our results with various benchmarks. Then, we investigate the effects of a dipole-dipole interaction, more closely related to experiments, and we show, for a realistic choice of parameters, that a supersolid cluster phase indeed occurs in shell geometry. Remarkably, this supersolid is resistant to the gravity of Earth.

To keep contact with the experiments, we simulate spinless bosons in three-dimensional (3D) space under the constraint of an external potential analogous to that realized in the lab. For $N$ particles with mass $m$, the Hamiltonian reads:
\begin{equation}\label{eq1}
H = \sum_{i=1}^{N} \left(-\lambda \nabla_{\textbf{r}_i}^2 + u_{\rm ext}(|\textbf{r}_i|) \right) + \sum_{i<j} v_{\rm int}(\textbf{r}_i-\textbf{r}_j),
\end{equation}
where $\textbf{r}_i$ is the position of the $i$-th particle, $\lambda = \hbar^2/2m$, $v_{\rm int}$ is the (possibly anisotropic) interaction potential, and
\be
u_{\rm ext}(d)=(u_0/\Omega)\sqrt{(d^2-\Delta)^2/4+\Omega^2}-u_0
\label{eq2}
\ee
is the external potential appropriate to a spherically-symmetric bubble trap~\cite{PhysRevA.98.013609} centered at the origin. In Eq.~(\ref{eq2}), $\Delta$ and $\Omega$ are square-length parameters related to the detuning and Rabi frequency of the rf field, respectively, while $u_0$ characterizes the harmonic trap prior to dressing. The potential (\ref{eq2}) interpolates between the filled sphere and the spherical shell (see Fig.~\ref{fig1}g). In the thin-shell limit, $u_{\rm ext}$ becomes harmonic around the minimum at $\sqrt{\Delta}$, which thus defines the radius $R$ of the reference sphere. As $R$ increases, the potential minimum gets more and more pronounced, until particles become effectively confined to a spherical surface.

In our PIMC simulations~\cite{cep95}, we employ the worm algorithm~\cite{PhysRevLett.96.070601} to sample the equilibrium statistics of bosons at finite temperature and estimate the superfluid fraction, $f_s$ \cite{supplementary}. To deal with the strong spatial constraint due to \eqref{eq2}, we introduce a biased version of the PIMC method~\cite{supplementary}. We also perform classical Monte Carlo (MC) and Molecular Dynamics (MD) simulations to probe the classical limit in the soft-core case.

We begin by considering bosons interacting through the soft-core potential:
\be
v_{\rm int}({\bf r})=\epsilon\theta(\sigma-r)\,,
\label{eq3}
\ee
where $\epsilon>0$ and $\sigma$ is the core diameter. In the discussion of the soft-core model, $\epsilon$ and $\sigma$ are taken as units of energy and length, respectively; temperatures are expressed in units of $\epsilon/{k_B}$. Due to the peculiar nature of the repulsion, at high density particles are gathered in droplets or clusters~\cite{likos01b}. Employing mean-field theory at $T=0$, the authors of Ref.~\cite{PhysRevA.99.063619} explored various possible arrangements of clusters on the sphere, finding the configuration of lowest enthalpy as a function of the radius. The evidence of supersolid phases is, however, not conclusive: condensation of clusters is assumed, not derived, while there is no guarantee that the true ground state has been identified. Using, for the first time, \textit{ab initio} PIMC simulations, we provide for the same interaction (\ref{eq3}) conclusive evidence of supersolid order. Moreover, we show how the supersolid withstands finite temperature in a realistic bubble trap, i.e., beyond a purely two-dimensional setup.

\begin{figure}[t]
  \begin{center}
    \includegraphics[width=\linewidth]{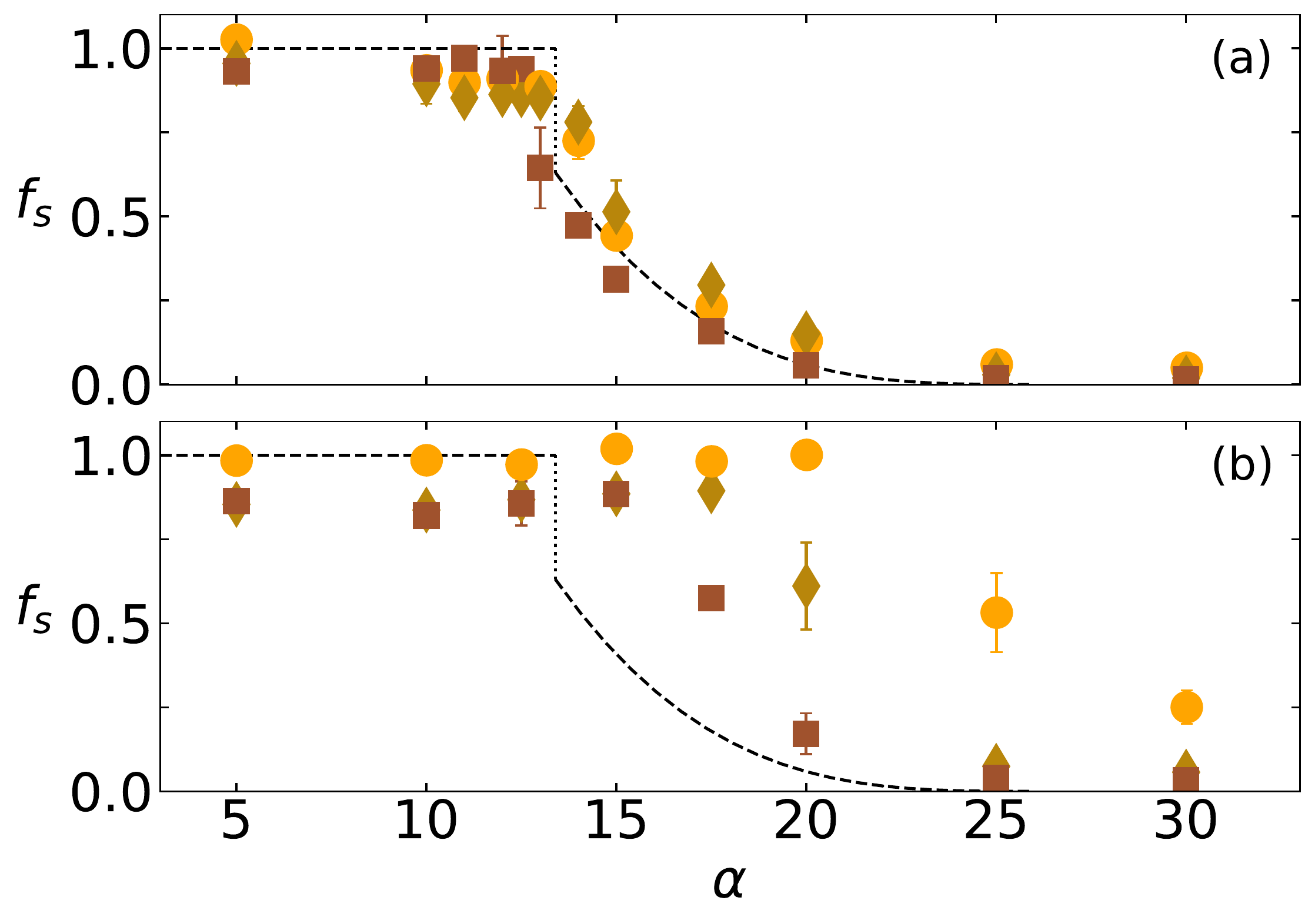}
    \caption{Superfluid fraction across the superfluid-supersolid-insulator transition, plotted as a function of $\alpha$ for $T=0.125$ (the $\rho$ entering the expression of $\alpha$ is $N/(4\pi R^2)$). (a) $u_0 = 50$. (b) $u_0 = 2$. Different symbols correspond to different system sizes: $N=40$ and $R=0.808$ (circles), $N=80$ and $R=1.143$ (diamonds), $N=120$ and $R=1.4$ (squares). The dashed lines refer to the planar limit~\cite{Macri:2014aa}.}
    \label{fig2}
  \end{center}
\end{figure}

First, we investigate the behavior of the system at a temperature $T \ll 1$, choosing $\Omega=0.0441$ (an arbitrary value much smaller than $\Delta$, see below). In Fig.~\ref{fig1}a-f we give a graphical account of the structures found for decreasing $\lambda$ at $T=0.125$, which are indicative of a superfluid-supersolid-insulator ``transition'', similarly as occurs on a plane~\cite{PhysRevA.87.061602}. Specifically, we perform a size scaling analysis of $f_s$ at fixed $N/(4\pi R^2)=4.87$, for two strengths of the trap potential, to see how the system approaches the planar limit. On a plane, the phases were characterized in terms of the dimensionless interaction strength $\alpha= m\rho\sigma^4\epsilon/\hbar^2 = \rho\sigma^4\epsilon / (2 \lambda) $ (for a number density $\rho=4.4$), finding that the superfluid-supersolid transition occurs at $\alpha\approx 13$, whereas the system becomes insulating at $\alpha\approx 22$~\cite{PhysRevA.87.061602}. In both two and three dimensions, the former transition is marked by a jump in $f_s$~\cite{PhysRevA.88.033618}. Our results are reported in Fig.~\ref{fig2}. In the strongly-confined case ($u_0=50$, top panel), the planar behavior is already recovered for $N=120$ particles, while for smaller sizes we find a smooth crossover. For a weaker external potential ($u_0=2$, bottom panel), the transition to supersolid is milder and shifted towards higher $\alpha$ values; moreover, convergence to the planar limit is much slower. In both cases, the system is a density wave at large $\alpha$; as $\alpha$ is reduced, quantum fluctuations increase, eventually leading to a disruption of polyhedral order which is faster when more freedom is given to particles in the radial direction, i.e., for $u_0=2$. The same sequence of transitions is reflected in the shape of the pair distribution function $g(r)$ (see Fig.~\ref{fig1}h and \cite{supplementary}). As long as $f_s>0$, $g(r)$ is non-zero at any distance, which is consistent with a condensate wave function being non-zero everywhere on the sphere. Instead, in a normal solid near zero temperature $g(r)$ would ideally be zero in the interstitial region between two successive shells of neighbors. The deviations from this behavior at moderate to large distances are a temperature effect, ultimately due to the finite extension of the clusters.

\begin{figure}[t]
  \begin{center}
    \includegraphics[width=\linewidth]{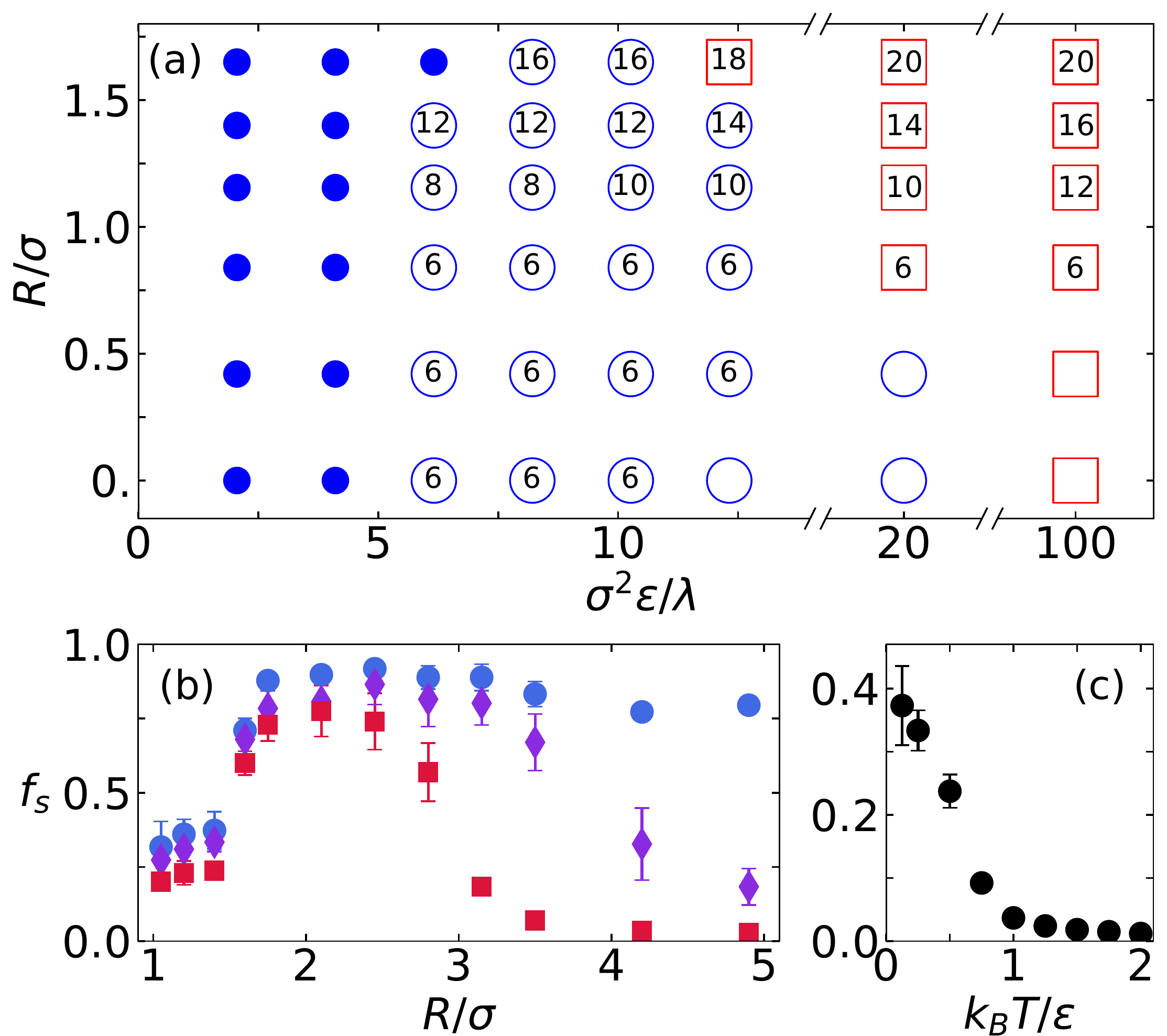}
    \caption{(a) A diagram showing, for $N=120$ and $T=0.5$, the number $N_c$ of clusters as a function of $\lambda^{-1}$ and $R$. The blue dots are state points where the system is superfluid. Open circles and squares mark supersolid and normal-solid states, respectively. When $\lambda$ and $R$ are both small, $N_c$ depends on the initial conditions (see more in \cite{supplementary}), therefore no number is printed in the symbol. (b) Superfluid fraction $f_s$ plotted as a function of $R$ at fixed $\lambda=0.16$ and $u_0=50$, for $T=0.5$ (red squares), 0.25 (purple diamonds), and 0.125 (blue circles). (c) $T$ dependence of $f_s$ for $R=1.4$ and $\lambda=0.16$.}
    \label{fig3}
  \end{center}
\end{figure}

Next, we fix the number $N$ of particles (120) and the temperature $T$ (0.5), and set $u_0=2$ and $\Omega=0.0441$ (we require $\Omega\ll 1$, so as to have $u_{\rm ext}(0)\approx\Delta/\Omega\gg 1$ for $R\approx 1$). We collect in a diagram the equilibrium arrangements for various $\lambda$ and $R$ (Fig.~\ref{fig3}a). On account of the $f_s$ value and of the evidence or not of polyhedral order, we can reasonably distinguish between superfluid, supersolid, and normal-solid states. Throughout the ``solid'' regions, the number $N_c$ of clusters may vary, but clusters are invariably found at the vertices of a regular or semiregular polyhedron: for example, an octahedron for $N_c=6$, a square antiprism for $N_c=8$, and an icosahedron for $N_c=12$~\cite{supplementary}.
For some $(\lambda,R)$ pairs, the cluster structure agrees with those predicted in \cite{PhysRevA.99.063619} (for example, the icosahedral structure at $\lambda=0.16$ and $R=1.4$). However, for several other pairs, the equilibrium configuration is a novel cluster phase not seen before. 
Moreover, contrary to the mean-field prediction, we see that decreasing $\/\lambda$ at fixed $R$ leads to transitions between equilibrium configurations with different numbers of clusters.
As we increase the radius, the supersolid region shrinks, while the number of clusters grows. At large radii, the particles assemble in smaller clusters, making it more difficult for the system to sustain superfluidity; eventually, above a certain $R$ the system is so dilute that clusters are washed away for every $\lambda$.

In the $T\to 0$ limit, the superfluid phase is stable at all large radii. Upon heating, similarly as in flat space, a superfluid-to-normal fluid transition eventually occurs. This is illustrated for $\lambda=0.16$ in Fig.~\ref{fig3}b, where we plot the superfluid fraction as a function of $R$ at different temperatures. As $T$ goes up, $f_s$ is gradually reduced until it vanishes, starting from larger $R$ values. At high enough temperatures, $f_s$ is also depleted for the supersolid, as we show in Fig.~\ref{fig3}c for $R=1.4$ and $\lambda=0.16$: interestingly, however, polyhedral order is preserved throughout the range of temperatures.

\begin{figure}[t!]
  \begin{center}
    \includegraphics[width=\linewidth]{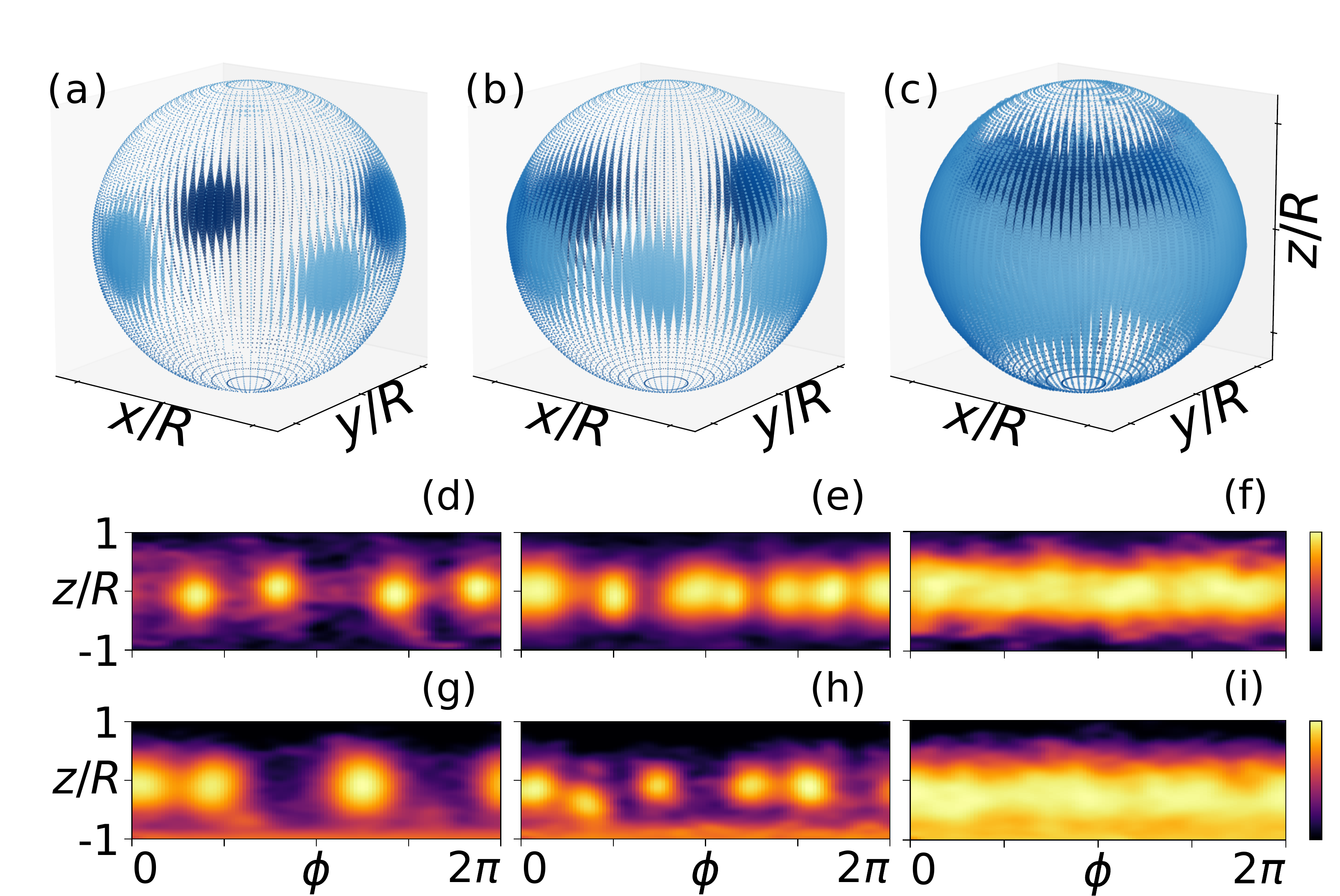}
    \caption{Dipolar atoms on a sphere. (a--c) Particle density for $N=360, T=1$ nK, and $R=2.6\,\mu$m, at $a/a_0=5$ (a), $a/a_0=50$ (b), and $a/a_0=350$ (c). The density increases with the size of the points, while colors are a guide for the eye. (d--f) Area-preserving projections of the densities in (a--c). (g--i) Effect of gravity. Brighter colors indicate larger values of the density (log scale).}
    \label{fig4}
  \end{center}
\end{figure}

When $\lambda \to 0$, the number of clusters approaches a value dependent on $R$; this regime corresponds to the classical limit, regardless of $R$ and $T>0$. To make sure that this is indeed the case, we have considered the classical counterpart of the quantum problem. The first observation of cluster phases in classical particles on a sphere goes back to Ref.~\cite{Franzini2018}, where density-functional-theory calculations are presented. Here, to keep the same level of accuracy of the quantum treatment, we carry out extensive MC and MD simulations of classical soft-core particles, using the same parameters of the quantum simulations. The results, expressed in terms of the number and final arrangement of clusters, are reported in \cite{supplementary} for various $R$ values. Except for small radii, where the uncertainty in $N_c$ is large, the aggregates formed by quantum particles have a comparatively smaller number of clusters. This is an effect of quantum delocalization, which causes the effective diameter of bosons to be larger than $\sigma$.

\begin{figure}[t!]
  \begin{center}
    \includegraphics[width=\linewidth]{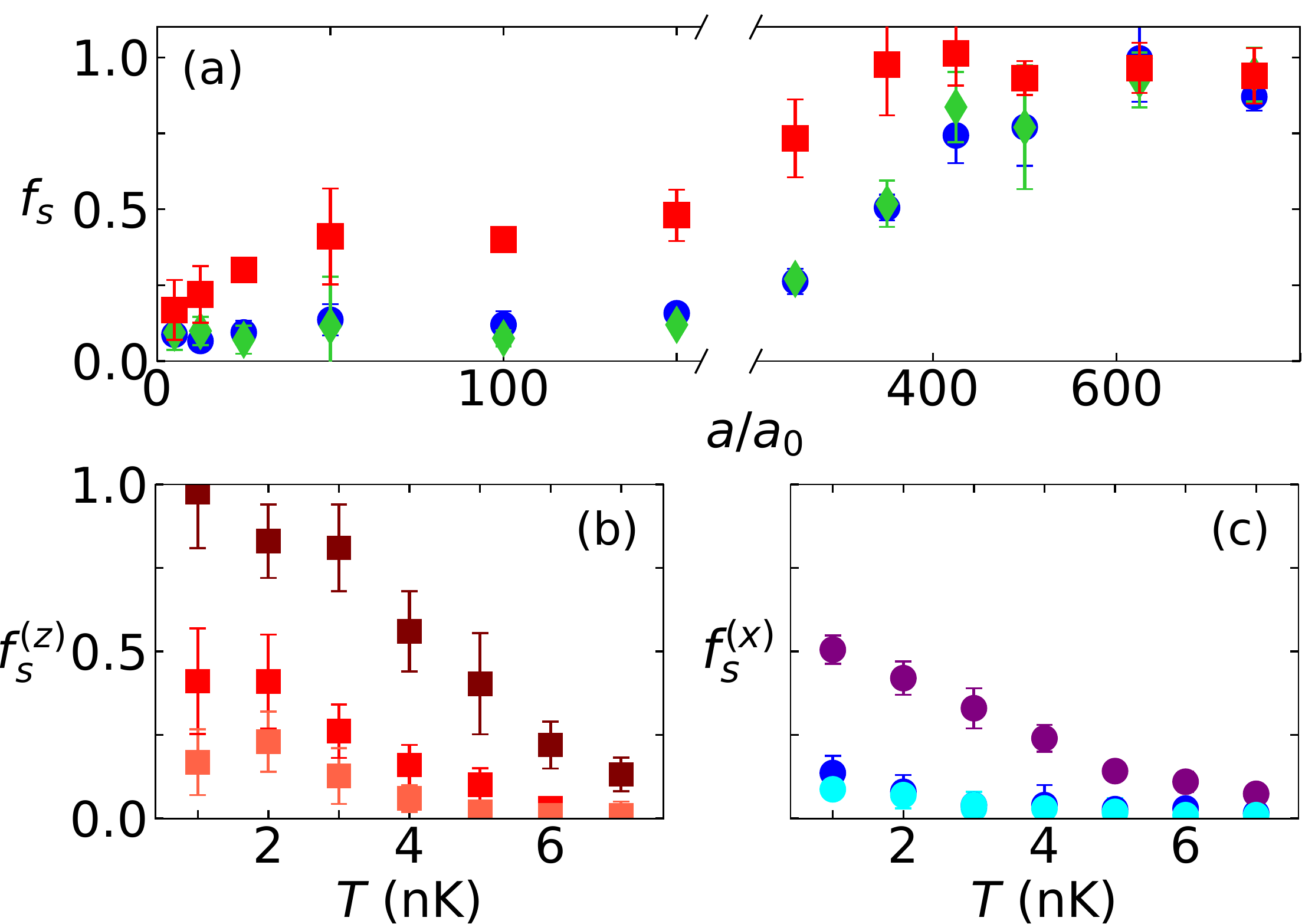}
    \caption{(a) Superfluid fractions $f_s^{(x)}$ (blue circles), $f_s^{(y)}$ (green diamonds), and $f_s^{(z)}$ (red squares) as a function of $a$. (b-c) Thermal behavior of $f_s^{(z)}$ and $f_s^{(x)}$, for $a/a_0=5$ (pink and light blue), 50 (red and blue), and 350 (brown and purple).}
    \label{fig5}
  \end{center}
\end{figure}

We now move to examine the behavior of dipolar bosons in the thin-shell limit. As is common in experiments, we assume atoms to be polarized along $\hat{\bf z}$. We work with a number density on the sphere of about $4\,\mu$m$^{-2}$, of the same order as that used in experimental realizations \cite{Lundblad2019}. The (anisotropic) interaction potential reads:
\begin{equation} \label{eq4}
v_{\rm int}(\textbf{r}) = v_{\rm HS}(r) + \frac{\mu_0 d_m^2}{4\pi}\frac{1-3\cos^2\theta}{r^3}\,,
\end{equation}
where $\mu_0$ is the vacuum permeability, $d_m= 9.93 \mu_B$ is the magnitude of the dipole moment of a $^{164}$Dy atom, $\mu_B$ is the Bohr magneton, and $\cos\theta=\hat{\bf r}\cdot\hat{\bf z}$ \cite{Baranov2008}.
Finally, $v_{\rm HS}(r)$ is the hard-sphere potential with core diameter equal to the scattering length $a$ \cite{supplementary}.

We simulate the system for various $a$, considering values up to $10^3 a_0$, with $a_0$ the Bohr radius, thus much smaller than the sphere radius, taken to be $R=2.6\,\mu$m (or $R \approx 50\times10^3 a_0$). Our results for $N=360$ and $T = 1$ nK are illustrated in Fig.~\ref{fig4}. 
As is clear from Eq.~(\ref{eq4}), particles attract each other along the $z$ direction and repel each other along $x$ and $y$. As long as $a$ is much smaller than $R$, particles move away from the poles of the sphere and bunch together around the equator. Indeed, up to $a\approx 150a_0$, particles form clusters lined up along the equator, as seen in Figs.~\ref{fig4}a and b. 
At variance with what observed for dipolar clusters in trapped Euclidean geometries \cite{Norcia2021}, the cluster phase on the sphere excludes zig-zag configurations: in Figs.~\ref{fig4}a and b, clusters remain on the equator due to the lack of a significant repulsion along $z$ that could counteract the mutual attraction.
Unless $a$ is very small, the superfluid fraction for rotations around $z$ is finite and significantly larger than $f_s$ for rotations around $x$ and $y$, thus qualifying this state as supersolid (Fig.~\ref{fig5}a) \cite{footnote1}.
For larger $a$, due to the increased particle exclusion, clusters grow in size and merge together forming a ribbon wrapped around the sphere, see Figs.~\ref{fig4}c and f. This specific configuration is consistent with the density profile seen in \cite{adhikari2012, Diniz2020}. Beyond $a\approx 200\,a_0$ the ribbon is homogeneous; however, between $a=150\,a_0$ and  $a=200\,a_0$, there is a wide crossover region where the clusters are still present as azimuthal density modulations. The modulated-to-unmodulated transition is shifted to slightly larger $a$ compared to other trapped geometries (see, e.g., \cite{sohmen21} and references cited therein) because of the enhanced stability of spherically-confined clusters. The superfluid fraction along $z$ is still significantly larger than in the other directions. Finally, for very large values of $a > 500a_0$, a homogeneous superfluid state emerges, similar to the one shown in Figs.~\ref{fig1}a and d. 
Simulations at different values of $N$ but fixed density show no significant scaling, a strong indication that the same supersolid phase persists at larger numbers of particles, up to experimental values.
The temperature analysis indicates that the supersolid behaves differently from the superfluid ribbon (see Figs.\,\ref{fig5}b and c). In the former case, $f_s^{(z)}$ remains non-zero up to $\approx 5$\,nK, due to a nonzero superfluid signal of the system around $z$. As for $f_s^{(x)}$, the signal is smaller and substantially constant, as a result of the thickness of the supersolid in the latitudinal direction. Instead, the ribbon remains superfluid up to much higher temperatures, while keeping its anisotropic character for all $T$.

Figs.~\ref{fig4}g--i depict simulations carried out by including the effect of gravity on $^{164}$Dy atoms. Interestingly, for small enough $a$ the atoms still form a ring of clusters perpendicular to $z$, although their center of mass is no longer at $z$=$0$, but slightly shifted downwards. For larger values of $a$, even though part of the atoms are found near the south pole, a shifted superfluid ribbon is still evident. This is to be contrasted with what happens when we remove the dipolar component of the interaction, but still keep the hard-core repulsion: in this case, all particles are pushed by gravity to the bottom of the trap~\cite{supplementary}.

To conclude, we have investigated the equilibrium phases of identical bosons in a shell trap. As the bubble expands, the system switches from three- to two-dimensional in a continuous fashion. 
In this peculiar setting, the PIMC algorithm needs an {\it ad hoc} modification that we have discussed at length in \cite{supplementary}. With this tool at hand, we have considered two species of bosons, i.e., soft-core atoms and dipolar atoms, providing evidence at small non-zero temperature of two unconventional supersolid behaviors. We argue that the realization of these phases is within reach of present-day technology. 
Very significantly, we have shown that the dipolar supersolid on the sphere is robust to the effects of gravity. This surprising result indicates that it could be probed experimentally even without sending the apparatus to space.
More generally, the ideas underlying our original implementation of the PIMC algorithm on the sphere can profitably be replicated for other curved surfaces, allowing to gain insight into the peculiarities of bosons in spaces where the geometric potential is non-zero (see more in the last section of \cite{supplementary}).

In this work, we have focused on demonstrating the existence of a supersolid phase on the sphere. Using the same numerical techniques, it would be also possible to investigate the nature of the superfluid phase on the sphere, e.g., BKT behavior or reentrant effects~\cite{sohmen21,SanchezBaena2023}.
Our work might also be relevant to supersolid phases in the crust of neutron stars~\cite{Recati2023}.
The dynamical properties of shell-trapped bosons are another open problem and an active area of research~\cite{PhysRevA.107.023319}. For example, it would be interesting to see whether the superfluid ribbon in Figs.~\ref{fig4}c and f can support persistent currents in the same way as does a superfluid ring~\cite{ryu2007,corman2014,guo20}.

\begin{acknowledgments}
The authors acknowledge the NICIS Centre for High-Perfomance Computing, South Africa, for providing computational resources. M. C. and F. C. acknowledge financial support from PNRR MUR Project No. PE0000023-NQSTI. Fruitful discussions with M. Boninsegni, T. Macrì, A. Mendoza-Coto, G. Modugno, and S. Moroni are gratefully acknowledged.
\end{acknowledgments}

\newpage
\clearpage

\setcounter{page}{1}

\section{Supplemental Material}

This document contains additional information on the quantum simulation method, including a detailed description of our biased PIMC method. Some more details on classical simulations are also given. Finally, in the framework provided by the Gross-Pitaevskii equation, we discuss the changes intervening in the Hamiltonian of a loosely constrained interacting system when the original 3D description is reformulated in a lower-dimensional curved manifold.

\subsection{Path integral Monte Carlo}

The application of the Path integral Monte Carlo (PIMC) method to particles moving near the surface of a sphere has prompted us to develop a specialized version of PIMC that we illustrate in this section. For ease of understanding, we first recall the foundations of PIMC and its most common implementations, with a focus on the proposal and acceptance of new configurations. 

Basically, PIMC relies on Feynman's path integral approach to the equilibrium statistical mechanics of quantum systems \cite{fey10,cep95,Boninsegni2005}. In this framework, all relevant information is contained in the canonical partition function, $Z$, which is the trace of the density matrix operator $e^{-\beta H}$, where $H$ is the Hamiltonian and $\beta=(k_BT)^{-1}$. In the canonical ensemble, the partition function of $N$ identical bosons reads:
\begin{equation}\label{eq:partf4}
Z = \frac{1}{N!} \sum_{P} \int d\textbf{R}\, \rho(\textbf{R},P\textbf{R},\beta),
\end{equation}
where
\begin{equation}\label{eq:partf2}
\rho(\textbf{R},\textbf{R}',\beta) = \braket{\textbf{R} | e^{-\beta H} | \textbf{R}'}
\end{equation}
is the density matrix and $P\textbf{R}$ = $(\textbf{r}_{P(1)},\textbf{r}_{P(2)},\ldots,\textbf{r}_{P(N)})$ is a permutation of particle coordinates.

The path integral approach consists in breaking up $\beta$ into $M$ (imaginary-)time slices of length $\tau = \beta /M$, and rewriting the partition function as a convolution of density matrices (viz., imaginary-time propagators) at a higher temperature:
\begin{equation}\label{eq:partf5}
\begin{split}
&Z=\frac{1}{N!}\sum_{P}\int d\textbf{R}^0d\textbf{R}^1\ldots d\textbf{R}^{M-1} \\ 
&\times \rho(\textbf{R}^0,\textbf{R}^1,\tau) \rho(\textbf{R}^1,\textbf{R}^2,\tau) \cdots\rho(\textbf{R}^{M-1},P\textbf{R}^0,\tau)\,,
\end{split}
\end{equation}
where $\textbf{R}^j=(\textbf{r}_{1}^{j},\textbf{r}_{2}^{j},\ldots,\textbf{r}_{N}^{j})$ is the vector of coordinates at a given time slice $j$. Therefore, associated with each particle $i$ are $M$ classical images $\textbf{r}_{i}^{j}$ (also called beads), forming its imaginary-time trajectory (also called a polymer, path, or worldline). Using approximate forms for $\rho$ at small $\tau$, the thermodynamic properties of the quantum system can be obtained by sampling suitable estimators on the Boltzmann distribution of a classical system of polymers, with the equivalence being exact in the limit $\tau\to 0$.

Worldline configurations can be sampled through a number of Monte Carlo schemes. The one we use, called the ``worm algorithm'', was developed in order to efficiently sample bosonic permutations; it works by opening the worldlines, allowing them to retract and extend in imaginary time \cite{PhysRevLett.96.070601}, and, most importantly, to join with each other. This may lead to the formation of  paths spanning a number of particles, which provides the visual representation of quantum coherence; sampling these paths properly is essential to correctly compute expectation values at low temperature.

Regardless of the scheme used, a crucial ingredient in the implementation of the PIMC method is the free-particle propagator. For simplicity, we specialize our discussion to the case of a single particle, but all concepts can be generalized to the many-particle case in a straightforward way. In the absence of external and interaction potentials, the density matrix for a single particle reads:
\begin{equation}
\label{eq:free_prop1}
\rho_{free} (\textbf{r}, \textbf{r}^\prime; \tau) = \frac{1}{(4\pi\lambda \tau)^{d/2}} \exp \left\{-\frac{(\textbf{r} -\textbf{r}')^2}{4\lambda \tau} \right\}\,,
\end{equation}
whereas, when potentials are present, in the limit $\tau \to 0$ it can be approximated as
\begin{equation}
\label{eq:free_prop2}
\rho(\textbf{r}, \textbf{r}^\prime; \tau) = \rho_{free} (\textbf{r}, \textbf{r}^\prime; \tau) \, e^{-\tau V(\textbf{r})} \,,
\end{equation}
called primitive approximation. In our simulations, we adopt this approximation to model the soft-core potential and the dipolar potential outside the core; the hard-core part of the latter is instead treated with the Cao-Berne propagator~\cite{caoberne, pil08, doi:10.7566/JPSJ.85.053001}. Convergence in $\tau$ is obtained for a number of timeslices between 64 and 256 in all cases. In a Monte Carlo scheme, new configurations are proposed by sampling a product of free-particle propagators and accepted or rejected based on the changes occurred in the external and interaction potentials. This is easily done computationally because the propagator \eqref{eq:free_prop1} between two successive polymer beads has the form of a Gaussian distribution with variance $\sigma = \sqrt{2 \lambda \tau}$, for which well-established sampling algorithms exist. In the context of the worm algorithm, which works with open worldlines, the simplest move is the propagation of the particle position from time slice $j$ to the next one. The displacement of position $\textbf{r}^{j+1}$ with respect to $\textbf{r}^{j}$ is drawn from the distribution $\rho_{free} (\textbf{r}^{j}, \textbf{r}^{j+1}; \tau)$ and accepted or rejected depending on change in the potential, $V(\textbf{r}^{j+1}) - V(\textbf{r}^{j})$. More generally, while updating the position of one bead at a time results in a slow simulation process, much more efficient moves are available: an essential ingredient of all PIMC algorithms are multilevel moves \cite{Boninsegni2005}, which update an entire portion of worldline, open or closed, at once. Finally, moves that shift worldlines rigidly in space are also routinely employed.

More sophisticated techniques exist which, by relying on better approximations to the potential term, build upon the above procedure to improve the efficiency of the acceptance step. When it comes to the external potential, a special case is represented by the harmonic potential: in this case, an analytic solution is available for the full propagator
\begin{equation}
\rho_{osc}(\textbf{r},\textbf{r}',\tau) = \braket{ \textbf{r} | \exp \left\{ -\tau \frac{\hat{\textbf{p}}^2}{2m} + \frac{m \omega^2 \hat{\textbf{r}}^2}{2} \right\} | \textbf{r}' } \, ,  
\end{equation}
requiring no separation between the kinetic and potential terms. In general, however, the external potential must be treated in the primitive approximation or through higher-order approximations~\cite{Chin1997}. As we now explain, this poses a problem for the bubble-trap potential, since, for $T\to 0$, it leads to reject almost every proposed configuration. 

The size of a free particle --- as well as the average diameter of the 3D polymer generated by sampling \eqref{eq:free_prop1} --- is proportional to the thermal wavelength, and therefore increases as the temperature is reduced, eventually becoming infinite when $T=0$. It is immediate to realize that, no matter how the confining potential is chosen, most polymer beads will eventually lie in regions of space that are energetically unfavorable. To avoid this, it is necessary to work with a very large number of timeslices and displace only a few beads at a time, so that the displacements are on the same scale of the characteristic length associated with the external potential; clearly, this would make the simulation very slow. As we have seen, in the case of a harmonic potential this problem can be circumvented by using a harmonic propagator in place of the free-particle propagator. However, this would not hold for arbitrary confining potentials; for this reason, the use of PIMC is usually limited to free or harmonically confined interacting systems, or to bounded external potentials such as those featuring optical lattices \cite{cia22}. 

\subsection{Biased PIMC method for bubble traps}

The biased PIMC algorithm proposed herein generates new configurations according to a distribution that is loosely tailored to fit the external potential. Compared to the standard algorithm, we introduce the following changes:

1) When attempting the propagation of a particle from time slice $j$ to time slice $j+1$, instead of generating the displacement from \eqref{eq:free_prop1}, we draw it from an anisotropic Gaussian distribution. A radial displacement is proposed with variance $\sigma_r$. An angular displacement is also proposed, derived from drawing a length from a two-dimensional Gaussian distribution with variance $\sigma_{\theta}$. This length must then be converted into an angle; there are different ways to do this, so the specifics of the move can vary, but this is not important since detailed balance will take care that small differences in the generated paths are smoothed out. For example, we can generate the length of arcs along the sphere; in this case, we draw $l_{\theta}$ from the Gaussian distribution and set $\theta = l_{\theta} / R$, leading to
\begin{equation}
    P(\Delta r, \Delta \theta) = \frac{1}{\sqrt{2\pi}\sigma_r} e^{-\frac{\Delta_r^2}{2\sigma_r^2}} \frac{1}{2\pi\sigma_{\theta}^2} e^{-\frac{\Delta_{\theta}^2 R^2}{2\sigma_{\theta}^2}}.
\end{equation}
Another option is that $l_\theta$ is the length of a chord, in which case $\theta= 2 \arcsin\left(l_\theta/2R\right)$. We have tested both solutions and found no significant differences in the results.

2) In a multilevel move, we construct successive positions along an arc rather than along a straight line.

3) Instead of proposing rigid shifts of a worldline in 3D space, we make the move in two steps: a radial shift first and then a rotation of the worldline around the center of the sphere.

A move is drawn from a proposal kernel $T(R\to R')$, then accepted with probability $W(R\to R')$. In order for detailed balance to be satisfied, we adopt the Metropolis rule
\begin{equation}
\label{rate_full}
W(R \to R') = \text{min}\left\{ 1, \frac{T(R'\to R)}{T(R\to R')} \frac{\pi(R')}{\pi(R)}\right\}\,,
\end{equation}
where $\pi$ is the equilibrium distribution.

\begin{figure}[t!]
    \centering
    \includegraphics[width=\linewidth]{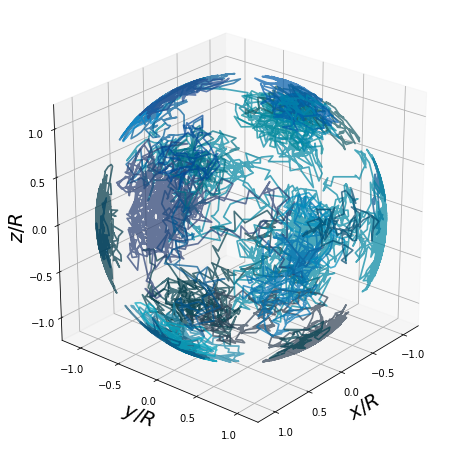}
    \includegraphics[width=\linewidth]{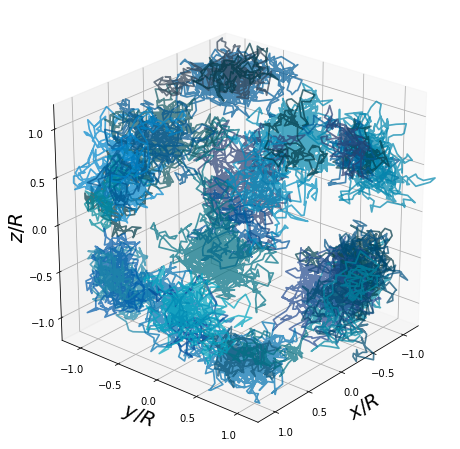}
    \caption{Snapshot of a typical configuration of the soft-core system in the polymer representation at $T=0.5$, $R=1.4$, and $\lambda = 0.16$. Each of the worldlines consists of $M=64$ beads. The shades of blue distinguish between distinct groups of connected worldlines. Top: $u_0=\infty$. Bottom: $u_0=2$. In either case the number of clusters is 12.}
    \label{fig_worldline}
\end{figure}

\begin{figure}[t!]
    \centering
    \includegraphics[width=\linewidth]{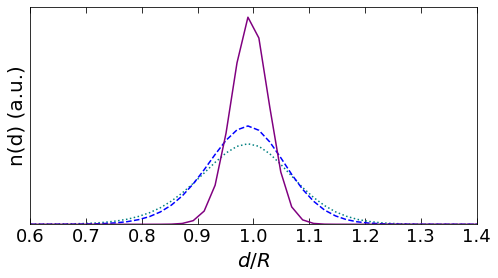}
    \caption{Radial density profile at $N=120,T=0.125,R=1.4$, and $\lambda=0.5$ (blue dotted line), 0.16 (blue dashed line), 0.01 (purple solid line). Here $u_0=2$.}
    \label{fig_radialdensity0}
\end{figure}

In the standard algorithm, $T(R \to R') \propto \rho_{free}(R')$, and since $\pi(R) \propto \rho_{free}(R') e^{-V(R')}$, the $\rho_{free}$ terms will cancel out, leading to 
\begin{equation}
\label{rate_free}
W(R \to R') = \text{min}\left\{ 1, \frac{e^{-V(R')}}{e^{-V(R)}}\right\}.
\end{equation}
In the biased algorithm, $T(R \to R')$ is a different distribution and all terms in \eqref{rate_full} must be computed explicitly. This may cause issues when $\sigma_r$ or $\sigma_\theta$ are very different from $\sqrt{2\lambda\tau}$, because large terms in the fraction can lead to either very large or very small acceptance rates depending on the move, especially in the worm algorithm when beads are removed altogether. For this reason, we employ symmetric moves where, whenever a worldline segment is deleted, it is immediately replaced by a new one. It is also possible to simulate the limiting case where $u_0 \to \infty$ by simply setting $\sigma_r=0$. The above procedure is applied for both closed worldline configurations and ``worm moves" involving open worldlines. As an illustrative example, in Fig.~\ref{fig_worldline} we show the worldline configuration at a single simulation step in equilibrium, corresponding to the cluster supersolid in Fig.~\ref{fig1}b.

To confirm that bosons are confined near the spherical surface, we look at the radial density profile, which is the particle density integrated over the solid angle. In Fig.~\ref{fig_radialdensity0}, we show the radial density at three values of $\lambda$, corresponding to the three cases of Fig.~\ref{fig1}. Indeed, we see that particles are lying near the spherical surface, as expected from the implementation of our biased PIMC for bubble traps. Interestingly, when the effects of delocalization are important (superfluid phase at $\lambda=0.5$, cyan dotted line; and supersolid phase at $\lambda=0.16$, blue dashed line), the distributions appear broader than in the other case of the normal cluster solid ($\lambda=0.01$, purple solid line), which instead displays a sharper peak. 

\begin{figure}[t!]
    \centering
    \includegraphics[width=\linewidth]{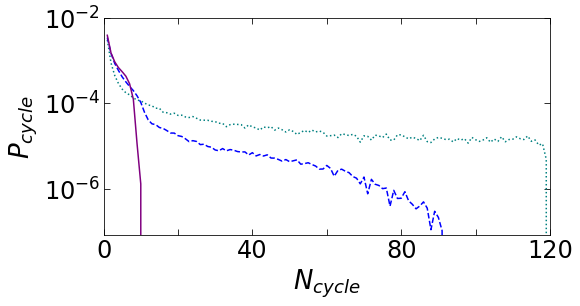}
    \caption{Relative frequency with which permutation cycles of various lengths occur at $N=120,T=0.125,R=1.4$, and $\lambda=0.5$ (blue dotted line), 0.16 (blue dashed line), 0.01 (purple solid line). Here $u_0=2$.}
    \label{permutations}
\end{figure} 
A key point to explore the effects of quantum statistics through a PIMC simulation is the distribution of polymer lengths, which is intrinsically connected to the indistinguishability of the particles and reflects quantum coherence as well as delocalization \cite{PhysRevB.84.014534}. In this framework, we consider the probability $P_{\rm cycle}$ to find polymer chains involving $N_{\rm cycle}$ particles, which we call permutation cycles ($1\leq N_{\rm cycle} \leq N$). This probability is obtained by updating at each step a histogram of $N_{\rm cycle}$. Fig.~\ref{permutations} reports results for the same set of parameters in Fig.~\ref{fig1}a-c. In the superfluid regime ($\lambda=0.5$, cyan dotted line) we find permutations extending over groups of particles of any number of particles. As the kinetic term decreases ($\lambda=0.16$, blue dashed line), the system enters a supersolid regime where we still observe permutation cycles over a considerably large amount of particles, again implying global coherence among clusters. Finally, for small $\lambda$ values ($\lambda=0.01$, purple solid line) permutations only involve the bosons contained in the same cluster. Here we have a normal solid phase, where quantum effects are limited to the single droplet.

\subsection{Superfluidity and area estimator}

A strong point of the PIMC method is the possibility to identify a phase with a periodic modulated density as supersolid \cite{RevModPhys.84.759}. Indeed, a characteristic feature of supersolids is their response to a slow axial rotation~\cite{Leggett1970}, which occurs with a reduced moment of inertia relative to an ordinary solid of same mass. The superfluid fraction $f_s$, i.e., the relative reduction in the moment of inertia, can be accurately estimated in a PIMC simulation.

In bubble traps, $f_s$ has been evaluated by sampling the so-called ``area estimator'' \cite{Sindzingre1989}. This method draws a direct connection between the area enclosed by tangled paths of polymers in a finite system and the reduction of the moment of inertia of the particles compared to the classical case. For reasons of symmetry, we only inspect the superfluid fraction along three orthogonal axes ($k=\,x,\,y$, and $z$). When doing so, the formula for $f_{s}^{(k)}$ reads:
\begin{equation}
\label{superf1}
f_{s}^{(k)} = \frac{4m^2}{\hbar^2}\beta I_{cl}^{(k)}\left(\langle A_{k}^2 \rangle - \langle A_{k}\rangle^2\right)\,,
 \end{equation}
where we keep the full definition by Sindzingre \textit{et al.} in Ref.~\cite{Sindzingre1989,cia22}.
In this formula, $A_{k}$ is the total area enclosed by particle paths projected onto the plane perpendicular to axis $k$, which can be written in terms of particle positions as 
\begin{equation}
\label{superf2}
A_k =  \frac{1}{2} \sum_{i=1}^{N} \sum_{j=0}^{M-1} \left(\textbf{r}_{i}^{j} \times \textbf{r}_{i}^{j+1} \right)_k\,,
\end{equation}
whereas $I_{cl}^{(k)}$ is the classical moment of inertia around the $k$ axis.

For the isotropic soft-core fluid, the superfluid fraction along any of the principal directions is the same within errors; therefore, we only report data for $f_s = (f_s^x + f_s^y +f_s^z)/3$. Conversely, in the dipolar case the anisotropy of the interaction makes it so that $f_s^z$ differs from $f_s^x$ and $f_s^y$ over a significant range of parameters; therefore, in figures their values are reported separately.

\subsection{The pair distribution function}

For clarity, here we describe the method of computing the pair distribution function, whose formal definition as a thermal average is (for spinless particles):
\begin{equation}
\label{rdfformula}
g(r)=\rho^{-2}\langle\Psi^\dagger({\bf x})\Psi^\dagger({\bf x}')\Psi({\bf x}')\Psi({\bf x})\rangle
\end{equation}
where $r=|{\bf x}-{\bf x}'|$ is the Euclidean distance in three dimensions, $\rho$ is the number density, and $\Psi$ and $\Psi^\dagger$ are field operators (an average over the direction of ${\bf x}-{\bf x}'$ is implied in (\ref{rdfformula})). In a PIMC simulation, $g(r)$ is sampled as a histogram. The interparticle distance $r$ is divided into bins, and a reference particle is chosen; at each sampling step, the distance between the reference particle and each other particle is calculated, and the value of the corresponding bin in the histogram is increased. The resulting histogram is then averaged over sampling steps. In Euclidean three-dimensional space, normalization is accomplished by dividing by a factor of $4\pi r^2$; in the present case, a number of different choices could be made, but we decide to keep the same normalization. In a PIMC simulation, sampling is done at each time slice and further averaged over time slices.

\subsection{Simulations with gravity}
\begin{figure}[t!]
    \centering
    \includegraphics[width=\linewidth]{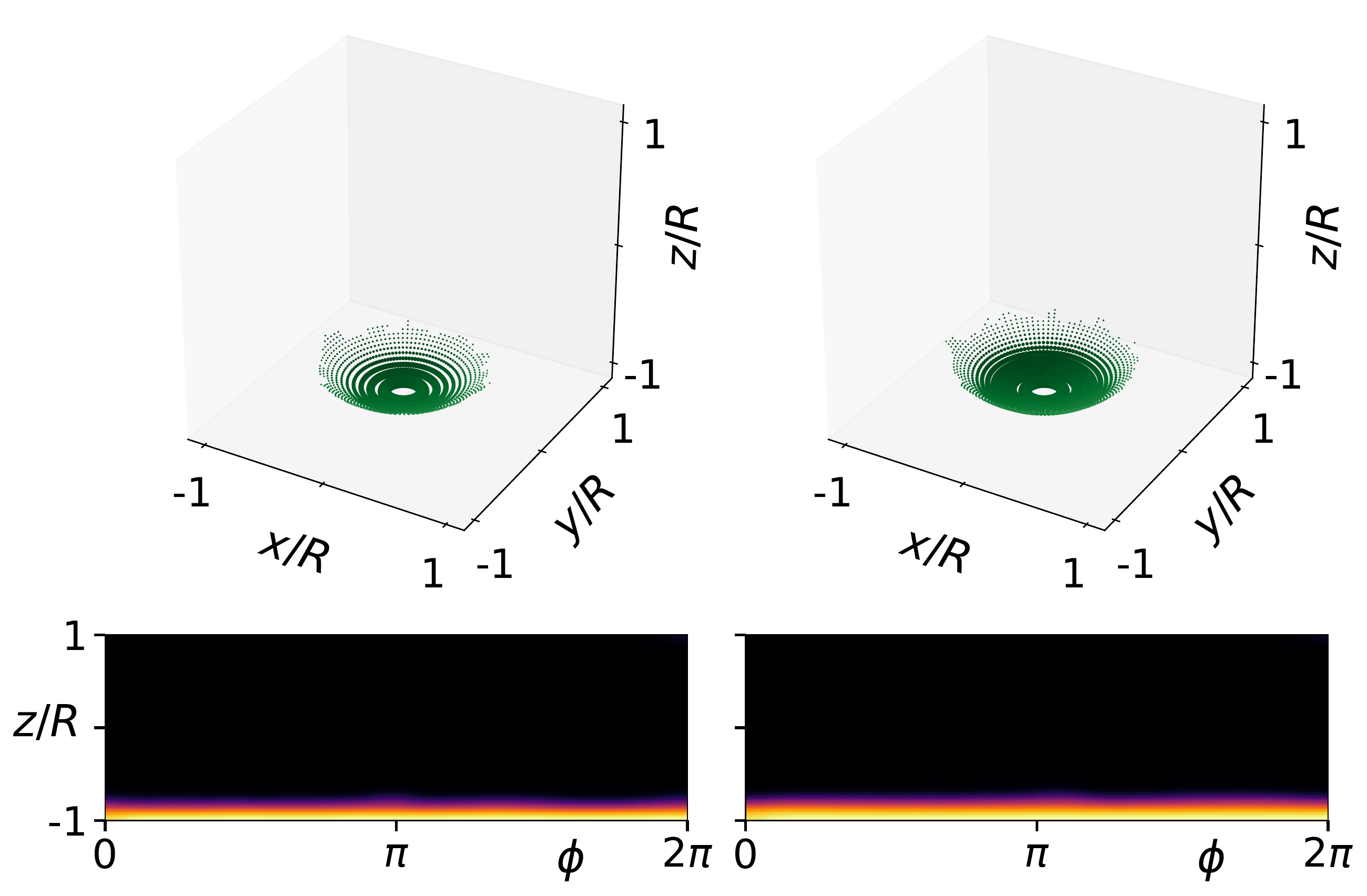}
    \caption{Non-interacting bosons (left) and bosons with hard-core interaction, $a=50 \,a_0$ (right) in the presence of gravity. In both cases, as expected, particles fall to the bottom of the trap.}
    \label{fig_sag}
\end{figure}

We have performed simulations in the presence of a gravitational field with $g=9.81 \, \text{m/s}^2$. In Fig~\ref{fig_sag}, we show results for the non-interacting and hard-core cases, confirming that the simulations give the expected result of particles sagging to the bottom of the bubble trap.

\subsection{Classical simulations}

We herein present, with more detail than provided in the main text, the results of Monte Carlo (MC) and molecular dynamics (MD) simulations of classical penetrable spheres of diameter $\sigma$, subject to the external potential in Eq.~(\ref{eq2}). For simplicity, in MD simulations we actually employ a continuous generalized exponential model potential with exponent 30, which would have practically the same behavior of the penetrable-sphere model. Simulations are performed at fixed $N=120,T=0.5,u_0=2$, and $\Omega=0.0441$, while the reference radius $R$ is varied through $\Delta$.

We employ the standard Metropolis MC algorithm with local and non-local moves. At each MC step, after randomly selecting a particle $i$ with position ${\bf r}_i$ and distance $d_i=|{\bf r}_i|$ from the trap center, a move is attempted from ${\bf r}_i$ to ${\bf r}_i'$, choosing between two equally likely possibilities: 1) ${\bf r}_i'$ is drawn at random from a cube of side $2\delta$ centered at ${\bf r}_i$; 2) ${\bf r}_i'$ is an arbitrary point on the sphere of radius $d_i$. As usual, $\delta$ is adjusted during the equilibration stage so as to ensure an average $40$-$50\%$ acceptance of type-1 moves. In MD simulations, the timestep is $dt=0.005 \sqrt{\frac{m \sigma^2}{ \epsilon}}$, where $m$ is the mass of the particles, and the thermal bath is modeled by a Nos\'e-Hoover thermostat with a temperature damping factor $\tau=100\,dt$ in time units. As for the initial configuration, we consider two possibilities: either the particles are placed at random on the reference sphere, or they are homogeneously distributed inside it (except for the case of the harmonic trap, where they are all placed at $R=0$). We generate hundreds of million MC steps and let the system evolve for a similar number of MD timesteps, to reduce the chance that the system might remain trapped in a metastable state. In the case of MD simulations, the initial velocities are randomly selected according to a Gaussian distribution. For $R$ less than 1.5, the system invariably relaxes to a clustered configuration. Once a stable cluster phase has formed, jump diffusion events~\cite{Moreno2007} between the clusters are extremely rare on the time scale of the simulations. As a function of $R$, we observe a sequence of ``transitions'' between different cluster arrangements (``phases''), which in MC simulations are all characterized by equally populated clusters, until the spherical density becomes so small that a homogeneous fluid is formed.

\begin{table}[h!]
\centering
\begin{tabular}{|c|c|c|} 
\hline
$R$ & classical & quantum ($\lambda=0.16$) \\ [0.5ex]
\hline\hline
0 & 5 (3+2), 7 (5+2) & 6 (OCT) \\
0.21 & 6 (OCT), 7 (5+2) & \\
0.42 & 4 (TET), 5 (3+2), 8 (SAP) & 6 (OCT) \\
0.63 & 4 (TET) & \\
0.84 & 6 (OCT) & 6 (OCT) \\
1.05 & 10 (SAP+2) & \\
1.155 & 12 (ICO) & 8 (SAP) \\
1.26 & 15 & \\
1.4 & 20 & 12 (ICO) \\ [0.5ex]
\hline
\end{tabular}
\caption{Number $N_c$ of clusters and their spatial arrangement as a function of $R$ (see text). The list of abbreviations is as follows. 4 (TET): 4 equivalent clusters at the vertices of a regular tetrahedron (TET). 5 (3+2): 3 clusters at the vertices of an equilateral triangle + 2 clusters on the axis of the triangle, farther from the origin. 6 (OCT): 6 equivalent clusters at the vertices of a regular octahedron (OCT). 7 (5+2): 5 clusters at the vertices of a regular pentagon + 2 clusters on the axis of the pentagon, closer to the origin. 10 (SAP+2): 8 equivalent clusters at the vertices of a square antiprism (SAP) with regular faces + 2 clusters on the axis of the antiprism, at the same distance from the origin. 12 (ICO): 12 equivalent clusters at the vertices of a regular icosahedron (ICO).}
\label{table1}
\end{table}

We report in Table 1 the final number $N_c$ of clusters and their overall arrangement for a few $R$ values, including for comparison also the available results of quantum simulations for $\lambda=0.16$. For the smaller $R$ values, two or three values of $N_c$ are quoted in the table, since the outcome of simulation showed a dependence on the simulation method and/or the initial configuration, suggesting the existence of several long-lived states that are close in free energy and are protected by high free-energy barriers. Similar problems exist in quantum simulations for small values of $\lambda$ and $R$. As a rule, relaxation to equilibrium is faster the larger is $R$.

To demonstrate the difference in relaxation time between small and large $R$ values, we have prepared two MD movies
~\cite{videoplaceholder}.
In the first movie~\cite{videoharmonic}, we show four segments from a MD simulation $5\times 10^8$ timesteps long, for the case $R=0$ (harmonic trap). Each segment of simulation lasts $5000$ timesteps and particle configurations are dumped every $10$ timesteps, corresponding to $500$ video frames. In the first segment, starting from $t=0$, we monitor the onset of a large cluster at the origin along with a few satellites made up of a single particle each. Following this event, for several hundreds of thousand MD timesteps we observe a system of four clusters, where the largest cluster still lies near the origin, while the remaining three clusters continue to grow. Only in the last few million MD steps, we observe the formation of a fifth cluster, initially consisting of a single particle, and we report this event in the second segment of the movie. In the third segment, we show the addition of a new particle to the fifth cluster, whereas in the last segment the addition of another particle makes the cluster finally populated with three particles. Our reasonable guess about the future evolution of the system is that the growth of the less-populated clusters will continue until the number of particles in all five clusters becomes nearly equal. In the second movie~\cite{videoicosahedron}, relative to $R=1.155$, we follow the simulation for $5000$ timesteps, starting from $t=0$, with particle configurations still dumped every $10$ timesteps. Starting from an initial configuration where particles are randomly distributed over the spherical surface, we monitor the quick formation of eleven clusters in the first hundred timesteps. Then, after approximately $3-4$ seconds of movie (about $1000$ timesteps), we see one of the clusters stretching and eventually breaking in two distinct clusters, thus giving rise to the stable icosahedral configuration referred to in Table 1. In summary, in the case of the harmonic trap ($R=0$), after $5\times 10^8 dt$ a structure with five clusters is established, but many more steps are likely needed in order that clusters become equally populated. On the contrary, for $R=1.155$, the approach to equilibrium is extremely fast: it only takes about $10^3 dt$ for the clusters to assemble in an icosahedral arrangement.

\begin{figure}[t]
\begin{center}
\includegraphics[width=9cm]{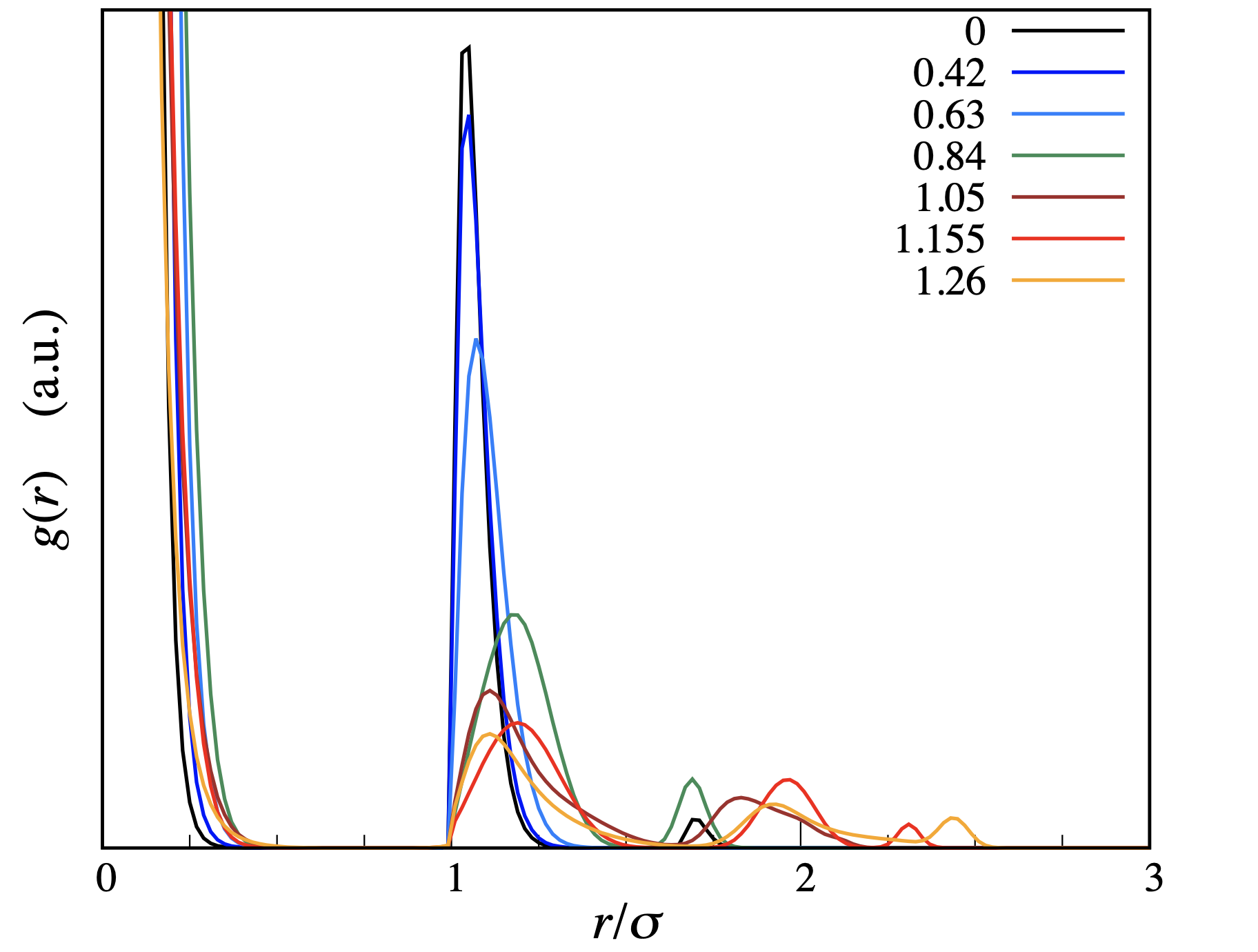}
\includegraphics[width=9cm]{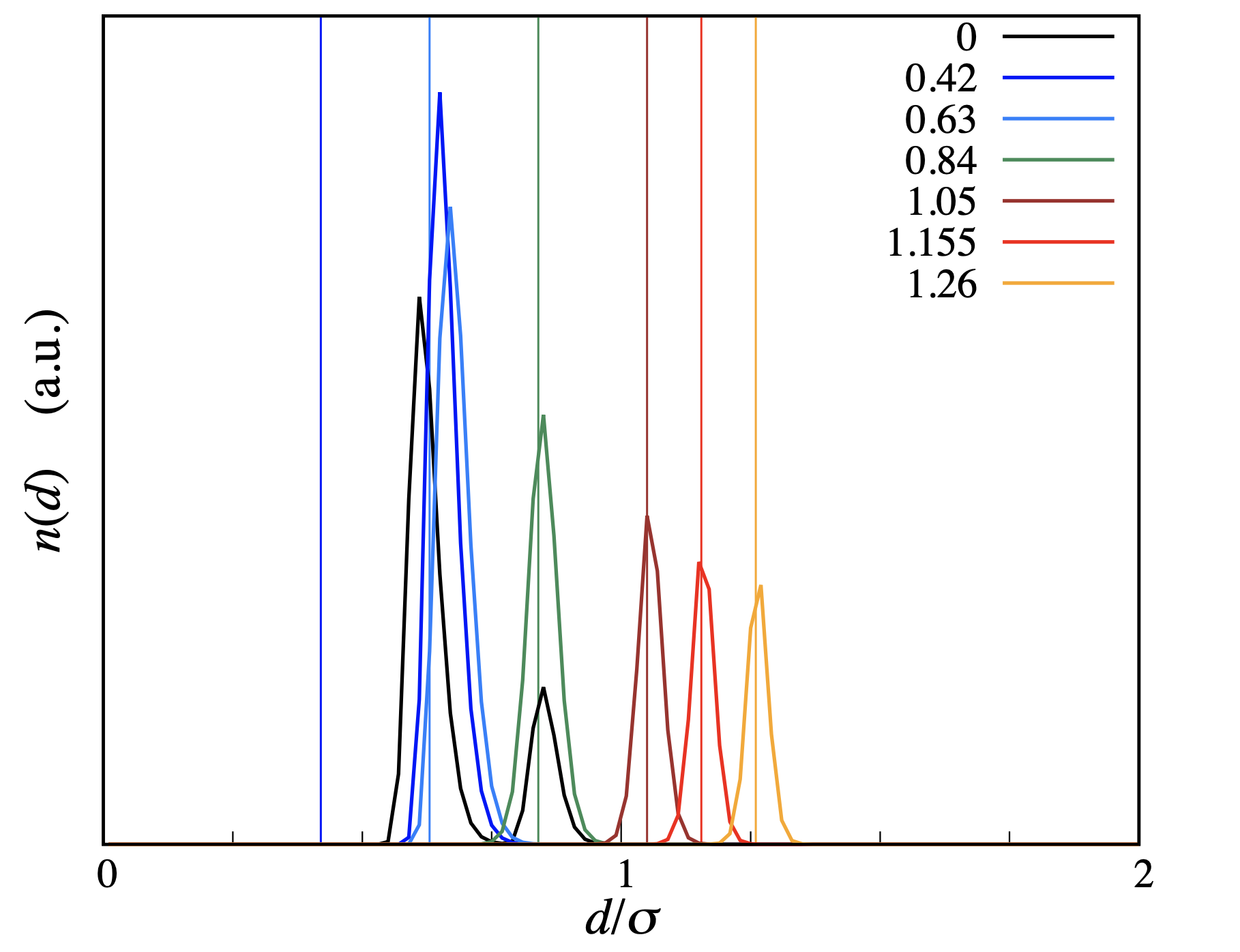}
\caption{Top: pair distribution function. Bottom: radial density. In the bottom panel, the vertical lines mark the various $R$ values. The only case where $n(d)$ shows two peaks is $R=0$ ($N_c=5$). Observe that for $R=0.42$ ($N_c=4$) the average distance of clusters from the origin is rather different from $R$. Also notice that $N_c=10$ for the $R=1.05$ case shown here.}
\label{sm2}
\end{center}
\end{figure}

We further illustrate the features of cluster phases by reporting in Fig.~\ref{sm2} results for two structural functions: the pair distribution function $g(r)$, $r$ being the Euclidean distance of a particle from a reference particle, and the radial density $n(d)$, where $d$ denotes the distance of a particle from the trap center. As $R$ grows, the transitions between different ``phases'' are reflected in qualitative changes of $g(r)$. The profile of $n(d)$ also changes, from one characterized by two peaks at $R=0$ to a single progressively sharper peak for $R>0$. For not too small radii, the rationale behind the choice of the equilibrium arrangement seems to be providing clusters with the most efficient occupation of the available space or, equivalently, with the highest coordination number $Z$. Due to geometric frustration, i.e., to the obstruction of triangular order due to curvature, this number cannot be six. As $R$ varies, the system adjusts the number of clusters so as to keep the distance between neighboring droplets always close to $\approx 1.25$ (see top panel of Fig.~\ref{sm2}); in turn, this implies an increase of $Z$ with $R$ towards the asymptotic value of 6.

\subsection{Thin-shell limit}

The quantization of a system of particles confined to a curved surface raises formidable challenges that can be pragmatically solved by regarding the constraint as the limit of a large restoring force arising when the particle coordinates deviate from their constrained values~\cite{Kaplan1997}. Atoms in a bubble trap provide a perfect representation of this viewpoint, as confinement to a spherical surface is here realized through an external potential that becomes progressively tighter as the radius $R$ of the reference sphere ${\cal S}$ increases. When $R$ is large enough the external potential reduces to a narrow harmonic well that constrains particles in the immediate neighborhood of the ${\cal S}$ surface. The question then arises as to what two-dimensional (2D) description is better suited to represent such a strong confinement. While simulation results remain the ultimate benchmark for the accuracy of any theoretical proposal, some clues on the nature of the reduced description are provided by zero-temperature calculations. Dimensional reduction at the level of the Gross-Pitaevskii (GP) equation has been considered in various papers: depending on the geometry of the trap, both one-dimensional (1D)~\cite{Leboeuf2001,Salasnich2002,Sandin2017,Salasnich2022} and 2D effective descriptions~\cite{Moller2020,Liang2022} can arise, where additional terms appear in the GP equation as a result of a strong (though not infinitely strong) confining field. Specifically, common to all such approaches is that the motion of quantum particles constrained to a curved surface is affected by a curvature-induced geometric potential~\cite{Jensen1971,DaCosta1981}. By elaborating further on the treatment put forth in \cite{Moller2020}, we sketch below the derivation of the generalized GP equation for an arbitrary interaction between the particles, and then specialize our discussion to the spherical surface.

Let $\Sigma$ be a regular surface and call ${\bf R}(u,v)$ the parametrization of a small piece of $\Sigma$ (``patch''). For points {\bf r} close to this portion of $\Sigma$, it is convenient to switch from Cartesian coordinates, ${\bf r}=(x,y,z)$, to new coordinates $q^i=(u,v,\zeta)$ (parallel and orthogonal to $\Sigma$), see e.g. Ref.~\cite{Prestipino2012}:
\be
{\bf r}={\bf R}(u,v)+\zeta\hat{\bf n}(u,v)\,,
\label{A1}
\ee
where
\be
\hat{\bf n}(u,v)=\frac{{\bf R}_u\wedge{\bf R}_v}{\left|{\bf R}_u\wedge{\bf R}_v\right|}
\label{A2}
\ee
is the unit normal to $\Sigma$. Transformation (\ref{A1}) is one-to-one sufficiently close to $\Sigma$. We will later consider the limit where the patch deviates only slightly from planarity, meaning that the local radii of curvature are much larger than the maximum allowed value of $|\zeta|$. Moreover, we denote $\Sigma(\zeta)$ the 2D manifold (``parallel'' to $\Sigma$) obtained by shifting all points of $\Sigma$ by $\zeta$ along the normal direction.

The metric tensor $G_{ij}$ (with $i,j=1,2,3$) relative to the transformation (\ref{A1}) reads:
\be
G_{ij}\equiv\frac{\partial{\bf r}}{\partial q^i}\cdot\frac{\partial{\bf r}}{\partial q^j}=
\begin{pmatrix}
g_{uu} & g_{uv} & 0\\
g_{vu} & g_{vv} & 0\\
0 & 0 & 1
\end{pmatrix}
\label{A3}
\ee
with
\ba
g_{\alpha\beta}&=&{\bf R}_\alpha\cdot{\bf R}_\beta+2{\bf R}_\alpha\cdot\frac{\partial\hat{\bf n}}{\partial q^{\beta}}\zeta+\frac{\partial\hat{\bf n}}{\partial q^{\alpha}}\cdot\frac{\partial\hat{\bf n}}{\partial q^{\beta}}\zeta^2
\nonumber \\
&\equiv&g_{\alpha\beta}^{(0)}+2s_{\alpha\beta}\zeta+h_{\alpha\beta}\zeta^2\,\,\,\,\,\,({\rm for}\,\,\alpha,\beta=1,2)\,.
\label{A4}
\ea
In differential geometry, $g_{\alpha\beta}^{(0)}$ and $s_{\alpha\beta}$ are respectively known as the first and second fundamental forms of $\Sigma$. The third fundamental form, $h_{\alpha\beta}$, is related to the previous two by~\cite{Hartman1953}
\be
h_{\alpha\beta}=-Kg_{\alpha\beta}^{(0)}+2Hs_{\alpha\beta}\,,
\label{A5}
\ee
$H$ and $K$ being the mean and Gaussian curvature of $\Sigma$, respectively.

As usual (see, e.g., Ref.~\cite{PhysRevA.99.063619}), the GP equation is derived by extremizing the grand potential per unit volume at $T=0$, $\omega[\psi]$, written as a functional of a trial condensate wave function $\psi$:
\ba
\omega[\psi]&=&\rho\int_{\cal N}{\rm d}^3r\,\psi^*({\bf r})\left[-\frac{\hbar^2}{2m}\nabla^2+V_{\rm ext}\right.
\nonumber \\
&+&\left.\frac{N-1}{2}\int_{\cal N}{\rm d}^3r'\,u(|{\bf r}-{\bf r}'|)|\psi({\bf r}')|^2-\mu\right]\psi({\bf r})\,,
\nonumber \\
\label{A6}
\ea
where $V_{\rm ext}({\bf r})$ is the confining field and ${\cal N}$ is the neighborhood of $\Sigma$ where particles are forced to lie. In order to separate the tangential from the normal motion, the above integral should first be written in terms of $(u,v,\zeta)$ coordinates (for each single patch of $\Sigma$). Then, the idea is to integrate out the $\zeta$ dependence everywhere, so as to remain with a functional of the 2D wave function only. After the change of coordinates, the kinetic-energy operator is still $-\hbar^2/(2m)$ times the Laplacian, for so it is prescribed by canonical quantization. Now, in curvilinear coordinates the Laplacian operator is given by
\be
\nabla^2=\frac{1}{\sqrt{\det G}}\frac{\partial}{\partial q^i}\left(\sqrt{\det G}\,G^{ij}\frac{\partial}{\partial q^j}\right)\,,
\label{A7}
\ee
as it follows from the Voss-Weyl formula for the divergence operator. Since $\det G=\det g$ and $G^{3j}=\delta_{j,3}$, we immediately find:
\be
\nabla^2=\nabla^2_\zeta+\frac{\partial\ln\sqrt{\det g}}{\partial\zeta}\frac{\partial}{\partial\zeta}+\frac{\partial^2}{\partial\zeta^2}\,,
\label{A8}
\ee
where
\be
\nabla^2_\zeta=\frac{1}{\sqrt{\det g}}\frac{\partial}{\partial q^\alpha}\left(\sqrt{\det g}\,g^{\alpha\beta}\frac{\partial}{\partial q^\beta}\right)
\label{A9}
\ee
is the Laplacian restricted to $\Sigma(\zeta)$.

Following the suggestion of \cite{Moller2020} we write the wave function of the condensate as
\be
\psi(u,v,\zeta)=\frac{1}{\sqrt[^4]{2\pi\sigma^2}}e^{-\frac{\zeta^2}{4\sigma^2}}\frac{\phi(u,v)}{\sqrt[^4]{\det g}}\,,
\label{A10}
\ee
where $\sigma$ (not to be confused with a particle diameter) is a small variational parameter and $\psi_r(u,v)=\phi(u,v)/\sqrt[^4]{\det g^{(0)}}$ represents the 2D wave function of the system. The Gaussian function in Eq.~(\ref{A10}) plays the role of system ground state along the normal direction, where the confining field acts as a harmonic potential:
\be
V_{\rm ext}(\zeta)=\frac{1}{2}m\omega^2\zeta^2\,.
\label{A11}
\ee
Assuming that the thickness $\sqrt{\hbar/(m\omega)}$ of ${\cal N}$ is much smaller than the minimum radius of curvature of the patch, the normalization of (\ref{A10}) should be effectively read as
\be
1=\int{\rm d}u\,{\rm d}v\,|\phi(u,v)|^2=\int{\rm d}u\,{\rm d}v\,\sqrt{\det g^{(0)}}|\psi_r(u,v)|^2\,.
\label{A12}
\ee

It is now straightforward, even though lengthy, to calculate the Laplacian (\ref{A8}) of the $\psi$ function in (\ref{A10}):
\ba
\nabla^2\psi&=&\left\{\left[\frac{\zeta^2}{4\sigma^4}-\frac{1}{2\sigma^2}-\frac{1}{4}\left(\frac{\partial\ln\sqrt{\det g}}{\partial\zeta}\right)^2\right.\right.
\nonumber \\
&-&\left.\frac{1}{2}\frac{\partial^2\ln\sqrt{\det g}}{\partial\zeta^2}-\frac{\zeta}{2\sigma^2}\frac{\partial\ln\sqrt{\det g}}{\partial\zeta}\right]\frac{\phi(u,v)}{\sqrt[^4]{\det g}}
\nonumber \\
&+&\left.\nabla^2_\zeta\frac{\phi(u,v)}{\sqrt[^4]{\det g}}\right\}\frac{e^{-\frac{\zeta^2}{4\sigma^2}}}{\sqrt[^4]{2\pi\sigma^2}}\,.
\label{A13}
\ea
The next step is to approximate the first and second derivatives of $\ln\sqrt{\det g}$ by their $\zeta=0$ values, and similarly replace the restricted Laplacian term with $\nabla^2_0\psi_r$, so that decoupling of coordinates becomes possible. A similar procedure yields for the interaction term:
\ba
&&\int_{\cal N}{\rm d}^3r\,\psi^*({\bf r})\int_{\cal N}{\rm d}^3r'\,u(|{\bf r}-{\bf r}'|)|\psi({\bf r}')|^2\psi({\bf r})
\nonumber \\
&=&\int{\rm d}u\,{\rm d}v\sqrt{\det g^{(0)}(u,v)}|\psi_r(u,v)|^2
\nonumber \\
&\times&\int{\rm d}u'{\rm d}v'\sqrt{\det g^{(0)}(u',v')}
\nonumber \\
&\times&\int{\rm d}\zeta{\rm d}\zeta'\,u(|{\bf r}-{\bf r}'|)\frac{1}{2\pi\sigma^2}e^{-\frac{\zeta^2+\zeta'^2}{2\sigma^2}}|\psi_r(u',v')|^2\,.
\nonumber \\
\label{A14}
\ea
The inner integral above is computed at the lowest order in $\sigma$, obtaining:
\be
\int{\rm d}\zeta{\rm d}\zeta'\,u(|{\bf r}-{\bf r}'|)\frac{1}{2\pi\sigma^2}e^{-\frac{\zeta^2+\zeta'^2}{2\sigma^2}}=w({\bf R}-{\bf R}',\hat{\bf n},\hat{\bf n}')\,,
\label{A15}
\ee
where
\ba
&&w({\bf x},\hat{\bf n},\hat{\bf n}')=u(x)+\sigma^2\left\{\frac{u'(x)}{x}\right.
\nonumber \\
&&-\left.\frac{1}{2}\left[\left(\hat{\bf n}\cdot\hat{\bf x}\right)^2+\left(\hat{\bf n}'\cdot\hat{\bf x}\right)^2\right]\left(\frac{u'(x)}{x}-u''(x)\right)\right\}
\nonumber \\
&&+{\cal O}(\sigma^4)\,.
\label{A16}
\ea
Hence, the curvature modifies the interaction between two particles, which acquires an additional dependence of order $\sigma^2$ on the normal vectors at the positions of the particles.

In conclusion, the grand potential becomes a functional of $\psi_r$:
\ba
&&\omega[\psi_r]\approx\rho\int{\rm d}u\,{\rm d}v\sqrt{\det g^{(0)}}\psi_r^*(u,v)\left\{-\frac{\hbar^2}{2m}\nabla^2_0\right.
\nonumber \\
&+&\frac{\hbar^2}{4m}\left[\frac{1}{2}\left(\frac{\partial\ln\sqrt{\det g}}{\partial\zeta}\right)^2+\frac{\partial^2\ln\sqrt{\det g}}{\partial\zeta^2}\right]_{\zeta=0}
\nonumber \\
&+&\frac{\hbar^2}{8m\sigma^2}+\frac{1}{2}m\omega^2\sigma^2+\frac{N-1}{2}
\nonumber \\
&\times&\int{\rm d}u'{\rm d}v'\sqrt{\det g^{(0)\prime}}\,w({\bf R}-{\bf R}',\hat{\bf n},\hat{\bf n}')|\psi_r(u',v')|^2
\nonumber \\
&-&\left.\mu\right\}\psi_r(u,v)\,.
\label{A17}
\ea
Upon imposing that the functional derivative of $\omega[\psi_r]$ with respect to $\psi_r^*$ is zero, we arrive at the GP equation in reduced dimensionality:
\ba
&&\mu\psi_r=\left(-\frac{\hbar^2}{2m}\nabla^2_0+V_{\rm q}+\frac{\hbar^2}{8m\sigma^2}+\frac{1}{2}m\omega^2\sigma^2+(N-1)\right.
\nonumber \\
&\times&\left.\int{\rm d}u'\,{\rm d}v'\sqrt{\det g^{(0)\prime}}\,w({\bf R}-{\bf R}',\hat{\bf n},\hat{\bf n}')|\psi_r(u',v')|^2\right)\psi_r\,,
\nonumber \\
\label{A18}
\ea
where
\be
V_{\rm q}(u,v)=\frac{\hbar^2}{4m}\left[\frac{1}{2}\left(\frac{\partial\ln\sqrt{\det g}}{\partial\zeta}\right)^2+\frac{\partial^2\ln\sqrt{\det g}}{\partial\zeta^2}\right]_{\zeta=0}
\label{A19}
\ee
is the {\em geometric potential}~\cite{Jensen1971,DaCosta1981} and $\sigma$ is to be adjusted so that the functional derivative of (\ref{A17}) with respect to $\sigma$ vanishes too (the optimal $\sigma$ would be of the order of $\sqrt{\hbar/(m\omega)}$). Clearly, Eq.~(\ref{A18}) resembles the standard GP equation, except for three additions: 1) a constant energy shift; 2) a ${\cal O}(\sigma^2)$ correction to the interaction between two particles; and 3) a curvature-induced external field $V_{\rm q}$.

For the spherical surface of radius $R$, the natural coordinates are the spherical angles $\theta$ and $\phi$. Then, $g_{\theta\theta}^{(0)}=R^2,g_{\theta\phi}^{(0)}=g_{\phi\theta}^{(0)}=0,g_{\phi\phi}^{(0)}=R^2\sin^2\theta$, while $H=1/R$ and $K=1/R^2$. Using the general formula~\cite{Moller2020}
\be
\det g=\det g^{(0)}(1+4H\zeta+(4H^2+2K)\zeta^2+4HK\zeta^3+K^2\zeta^4)\,,
\label{A20}
\ee
we obtain
\be
\det g=R^4\sin^2\theta\left(1+\frac{\zeta}{R}\right)^4\,.
\label{A21}
\ee
It then follows from the latter equation that the geometric potential (\ref{A19}) is identically zero for the spherical surface~\cite{Moller2020}.

\begin{thebibliography}{91}
\bibitem [{vid()}]{videoplaceholder}%
  \BibitemOpen
  \href@noop {} {\bibinfo {title} {The videos are available on request upon
  contacting the authors.}}\BibitemShut {Stop}%
\end{thebibliography}
\end{document}